\documentstyle[11pt]{article}\textheight 230mm\textwidth 150mm
\newcommand{\be}{\begin{eqnarray}}
\newcommand{\ee}{\end{eqnarray}}
\newcommand{\eel}[1]{\label{#1}\end{eqnarray}}
\newcommand{\e}[1]{\label{e:#1}\end{eqnarray}}
\newcommand{\r}[1]{(\ref{e:#1})}
\newcommand{\ben}{\begin{enumerate}}
\newcommand{\een}{\end{enumerate}}
\newcommand{\beq}{\begin{quote}}
\newcommand{\eq}{\end{quote}}

\newcommand{\vb}{{\cal h}}
\newcommand{\hb}{{\cal i}}

\newcommand{\ra}{{\rightarrow}}
\newcommand{\nn}{\nonumber}
\newcommand{\eg}{{\em e.g.\ }}
\newcommand{\ie}{{\em i.e.\ }}
\newcommand{\al}{\alpha}
\newcommand{\ga}{{\gamma}}
\newcommand{\la}{{\lambda}}

\newcommand{\del}{{\delta}}

\newcommand{\pet}{{\cal P}}
\newcommand{\bata}{\bar{\eta}}
\newcommand{\bapet}{\bar{\pet}}

\newcommand{\bett}{{\bf 1}}
\newcommand{\halv}{\frac{1}{2}}

\begin{document}
\begin{titlepage}
\noindent
G\"{o}teborg ITP 94-39\\
December 1994\\
hep-th/9501003\\
\vspace*{30 mm}
\begin{center}{\LARGE\bf BRST quantization of gauge theories\\ like SL(2,R)
 on inner product
spaces}\end{center}\begin{center} \vspace*{12 mm}

\begin{center}Robert Marnelius\footnote{tferm@fy.chalmers.se} and Ulrich
Quaade\footnote{Permanent
address: Odense Univ., Campusv 55, DK-5230 Odense M. E-mail: ujq@dou.dk}\\
\vspace*{7 mm}
{\sl Institute of Theoretical Physics\\
Chalmers University of Technology\\
S-412 96  G\"{o}teborg, Sweden}\end{center}
\vspace*{25 mm}
\begin{abstract}
Some general formulas are derived for the solutions of a BRST quantization
 on inner product
spaces of finite dimensional bosonic gauge theories invariant under
arbitrary Lie groups.
A detailed analysis is then performed of SL(2,R) invariant models and
some possible geometries of the
Lagrange multipliers are derived together with explicit results for a
class of SL(2,R) models. Gauge
models invariant under a nonunimodular gauge group are also studied in some
detail.
\end{abstract}\end{center}\end{titlepage}

\setcounter{page}{1}

\setcounter{equation}{0}
        \section{Introduction.}
In ref.\cite{Simple} it was shown  that a BRST quantization of a bosonic
gauge theory with an arbitrary Lie group invariance may be explicitly
solved on an inner product
space for finite degrees of freedom. The solutions were shown to have the form
\be
&&|ph\hb=e^{[\rho,Q]_+}|\Phi\hb
\e{1}
where $Q$ is the BRST charge operator and $\rho$ a fermionic gauge
fixing operator. $|\Phi\hb$ is a
state which does not depend on ghosts or Lagrange multipliers. Eq.\r{1}
is a formal solution which
actually may be extended to gauge theories of arbitrary ranks \cite{BM}.
In order to find out whether
or not it exists one must prescribe exactly the quantization properties
of the involved variables.
In \cite{Propa} some quantization rules were proposed which were shown
to be necessary in order for
states of the form \r{1} to have finite norms. These rules determine
the quantum properties of
the original inner product space with indefinite metric states.

In this paper we derive some general formulas for the solutions \r{1}
which are to be used in
conjunction with the quantization rules of \cite{Propa}. In particular
we  perform a
detailed analysis of general  gauge models whose gauge group is
SL(2,R) and determine some possible geometries of the Lagrange multipliers.
The results for an
explicit class of models are also derived. Finally we treat gauge models
invariant under a simple
nonunimodular gauge group.

\setcounter{equation}{0}
        \section{General properties.}
Consider a gauge theory in which the gauge generators $\psi_a,\;a=1,
\ldots,m$ are bosonic and
satisfy the Lie algebra
\be
&[\psi_a, \psi_b]_{-}=iU_{ab}^{\;\;c}\psi_c
\e{2}
where $U_{ab}^{\;\;c}$ are the structure constants. The BFV form of
the BRST charge operator is then
\cite{BV}
\be
&&Q=\psi_a\eta^a-\frac{1}{2}iU_{bc}^{\;\;a}\pet_a
\eta^b\eta^c-\frac{1}{2}iU_{ab}^{\;\;b}\eta^a + \bapet_a\pi^a
\e{3}
where $\eta^a$, $\bar{\eta}_a$ are Faddeev-Popov ghosts and antighosts
respectively and $\pet_a$,
$\bapet^a$ their conjugate momenta. $\pi_a$ are conjugate momenta
to the Lagrange multipliers $v^a$.
They satisfy the algebra (the nonzero part)
\be
&&[\eta^a, \pet_b]_+=[\bata^a,\bapet_b]_+=\del^a_b,\;\;\;[v^a,
\pi_b]_-=i\del^a_b
\e{4}
We assume that $\psi_a$ as well as all ghosts and Lagrange multipliers
are hermitian which is no
restriction. This implies that $Q$ in \r{3} is  hermitian. In \cite{Simple} it
was
shown that the general $Q$ given in \r{3} may always be decomposed as
\be
&&Q=\del+\del^{\dag}
\e{5}
where
\be
&&\del^2=\del^{\dag 2}=[\del, \del^{\dag}]_+=0
\e{6}
Based on the results of \cite{Bi} it was also shown that  the BRST
condition $Q|ph\hb=0$, which
projects out the physical states $|ph\hb$, may be solved by a bigrading which
requires
\be
&&\del|ph\hb=0,\;\;\;\del^{\dag}|ph\hb=0
\e{7}
The decomposition \r{5} is not unique. However, for the class of decompositions
found in  \cite{Simple,Gauge} the solutions of \r{7} could be written as
\be
&&|ph\hb_{\al}=e^{\al[\rho, Q]_+}|\Phi\hb
\e{8}
where $\al$ is a real constant different from zero, and $\rho$ a hermitian
fermionic gauge fixing
operator whose possible forms were derived. For the decompositions found
in \cite{Simple} it is of the
form
 \be
&&\rho=\pet_av^a
\e{9}
and where $|\Phi\hb$ satisfies the conditions
\be
&&\eta^a|\Phi\hb=\bar{\eta}_a|\Phi\hb=\pi_a|\Phi\hb=0
\e{10}
which simply means that $|\Phi\hb$ has no ghost dependence and no
dependence on the Lagrange
multipliers $v^a$. Thus, $|\Phi\hb$ only depends on the matter
variables. Since \r{10} is always
trivially solved, the expression \r{8} represents general
solutions for any gauge theory with finite
number of degrees of freedom and in which the gauge group is a Lie group.
(In \cite{BM} the solution
\r{8} is extended to any gauge group.) For the decompositions \r{5}
found in \cite{Gauge} there are
also formal solutions of the form  \r{8}, however, in this case $\rho$
involves gauge fixing
conditions to $\psi_a$ and $|\Phi\hb$ must be a solution of a Dirac
quantization
($\sim\psi_a|\Phi\hb=0$). Although these are also possible solutions
we concentrate on the solutions
\r{8} where $\rho$  and $|\Phi\hb$ satisfy \r{9} and \r{10}. They are of
a more geometrical nature
since $\rho$ is simple and $|\Phi\hb$ trivial.

The inner products of the solutions \r{8} have the form
\be
&&_{\al}\vb ph|ph\hb_{\al}=\vb\Phi|e^{2\al[\rho, Q]_+}|\Phi\hb
\e{11}
{}From \r{10} it follows that this expression can only be well
defined if $\al\neq0$ ($|\Phi\hb$ does
not belong to an inner product space). However, one may easily
prove that \r{11} is independent of the
value of $\al$ as long as it is positive or negative \cite{Simple,BM}.
In fact, we have for
the gauge fixing \r{9} \cite{BM} \be
&&|ph\hb_{\al}=U(\beta)|ph\hb_{\pm},\;\;\;|ph\hb_{\pm}\equiv|ph\hb_{\pm1}
\e{12}
where $\al=\pm e^{\beta}$ and where $U(\beta)$ is the unitary operator
\be
&&U(\beta)\equiv e^{-i\beta[v^a\bar{\eta}_a, Q]_+}
\e{13}
Eq.\r{12} means  that \r{11}  may only depend on the sign of $\al$.
As in \cite{Propa} it is natural
to impose the condition that \r{11} is independent of the sign of $\al$.
However, this is a nontrivial
condition. Notice \eg that there cannot be solutions $|ph\hb_{\al}$
for both positive and negative
$\al$'s at the same time if we have performed a projection from an original
inner product space
since  $_{-\al}\vb ph|ph\hb_{\al}=\vb\Phi|\Phi\hb$  is undefined. In fact,
as was clarified in
\cite{Propa}, the original state spaces yielding the solutions with
positive and negative $\al$ have
different bases. The condition that solutions with opposite signs of
$\al$ should yield equivalent
results requires a rather precise general condition on the ranges of
the Lagrange multipliers (see
below and \cite{Propa}).

Consider now the inner product \r{11} for the gauge fixing \r{9} with
$\al=\halv$. We
have
\be
&&[\rho, Q]_+=\psi_av^a+\psi_a^{gh}v^a+i\pet_a\bapet^a,\;\;\;\psi_a^{gh}\equiv
\halv
i{U_{ab}}^c(\pet_c\eta^b-\eta^b\pet_c)
\e{14}
In appendix A the following two equivalent expressions are derived
 \be
&&_{+}\vb ph|ph\hb_{+}=\vb\Phi|e^{[\rho,
Q]_+}|\Phi\hb=\left\{\begin{array}{c}\vb\Phi|e^{\psi'_av^a}
e^{i{(L^{-1})^a}_b(iv)\pet_a\bapet^b}|\Phi\hb\\
\vb\Phi|e^{i{(L^{\dag-1})^a}_b(iv)\pet_a\bapet^b}e^{\psi''_av^a}|\Phi\hb
\end{array}\right.
 \e{15}
where ${L^a}_b(iv)$ is the left invariant vielbein on the group manifold with
imaginary coordinates
$iv^a$ (${L^{\dag a}}_b(iv)={L^a}_b(-iv)$) (see also \cite{Simple}), and where
\be
&&\psi'_a\equiv\psi_a-\halv i{U_{ab}}^b,\;\;\;\psi''_a\equiv\psi_a+\halv
i{U_{ab}}^b=(\psi'_a)^{\dag}
\e{16}
Notice that
\be
&&e^{i{(L^{\dag-1})^a}_b(iv)\pet_a\bapet^b}e^{\psi''_av^a}=
(e^{\psi'_av^a}e^{i{(L^{-1})^a}_b(iv)\pet_a\bapet^b})^{\dag}
\e{17}

For $\al=-\halv$ in \r{11} we find in appendix A the expressions
 \be
&&_{-}\vb ph|ph\hb_{-}=\vb\Phi|e^{-[\rho,
Q]_+}|\Phi\hb=\left\{\begin{array}{c}\vb\Phi|e^{-\psi'_av^a}
e^{-i{(L^{\dag-1})^a}_b(iv)\pet_a\bapet^b}|\Phi\hb\\
\vb\Phi|e^{-i{(L^{-1})^a}_b(iv)\pet_a\bapet^b}e^{-\psi''_av^a}|\Phi\hb
\end{array}\right.
 \e{18}
where
\be
&&e^{-i{(L^{-1})^a}_b(iv)\pet_a\bapet^b}e^{-\psi''_av^a}=
(e^{-\psi'_av^a}e^{-i{(L^{\dag-1})^a}_b(iv)\pet_a\bapet^b})^{\dag}
\e{19}
Now we have
\be
&&e^{i{M^a}_b\pet_a\bapet^b}=(\det {M^a}_b)(i\pet_a\bapet^a)^m+\mbox{lower
 powers of $\pet_a$
and $\bapet^a$}
\e{20}
and
\be
&&|\Phi\hb=|\phi\hb|0\hb_{\pi}|0\hb_{\eta\bar{\eta}}
\e{21}
where $|0\hb_{\pi}$ is a Lagrange multiplier vacuum ($\pi_a|0\hb=0$),
$|0\hb_{\eta\bar{\eta}}$
 a ghost vacuum
($\eta^a|0\hb_{\eta\bar{\eta}}=\bar{\eta}_a|0\hb_{\eta\bar{\eta}}=0$)
 and $|\phi\hb$
an arbitrary matter state  which is spanned by the matter variables
involved in \eg $\psi_a$. (The physical part of $|\phi\hb$ should belong to an
inner product
space.) The ghost vacuum may be normalized as follows \cite{Fermi} \be
&&_{\eta\bar{\eta}}\vb 0|(i\pet_a\bapet^a)^m|0\hb_{\eta\bar{\eta}}=1
\e{22}
Making use of \r{20} and \r{21} in \r{15} and \r{18} we find
 \be
&&_{+}\vb
ph|ph\hb_{+}=\left\{\begin{array}{c}\vb\phi|\,_{\pi}\vb0|e^{\psi'_av^a}
\det{(L^{-1})^a}_b(iv)|0\hb_{\pi}|\phi\hb\\
\vb\phi|\,_{\pi}\vb0|\det{(L^{\dag-1})^a}_b(iv)e^{\psi''_av^a}|0\hb_{\pi}
|\phi\hb \end{array}\right.
 \e{23}
and
\be
&&_{-}\vb
ph|ph\hb_{-}=\left\{\begin{array}{c}(-1)^m\vb\phi|\,_{\pi}\vb0|e^{-\psi'_av^a}
\det{(L^{\dag -1})^a}_b(iv)|0\hb_{\pi}|\phi\hb\\
(-1)^m\vb\phi|\,_{\pi}\vb0|
\det{(L^{-1})^a}_b(iv)e^{-\psi''_av^a}|0\hb_{\pi}|\phi\hb
\end{array}\right.
 \e{24}

The Lagrange multipliers $v^a$ are hermitian operators. They have therefore
either real or imaginary
eigenvalues depending on whether or not they span a positive or an indefinite
metric
state space. In the special case when all $v$'s span positive metric states,
the eigenstates may be
normalized as follows
\be
&&\vb v|v'\hb=\del^m(v-v'),\;\;\;\int d^mv|v\hb\vb v|=\bett
\e{25}
Choosing the normalization of $|0\hb_{\pi}$ such that
\be
&&_{\pi}\vb 0|v\hb=\vb v|0\hb_{\pi}=1
\e{26}
we find for \r{23} and \r{24}
 \be
&&_{+}\vb
ph|ph\hb_{+}=\left\{\begin{array}{c}\int d^mv\det{(L^{-1})^a}_b(iv)\vb\phi
|e^{\psi'_av^a}
|\phi\hb\\
\int d^mv \det{(L^{\dag-1})^a}_b(iv)\vb\phi|e^{\psi''_av^a}|\phi\hb
\end{array}\right.
 \e{27}
\be
&&_{-}\vb
ph|ph\hb_{-}=\left\{\begin{array}{c}(-1)^m\int d^mv\det{(L^{-1})^a}_b(-iv)
\vb\phi|e^{-\psi'_av^a}
|\phi\hb\\
(-1)^m\int d^mv \det{(L^{\dag-1})^a}_b(-iv)\vb\phi|e^{-\psi''_av^a}|\phi\hb
\end{array}\right.
 \e{28}
In this case it is clear that convergence of these expressions will in general
 severely restrict the
ranges of $v^a$.

In the other extreme case when all Lagrange multipliers span indefinite
metric state spaces, $v^a$
has imaginary eigenvalues $iu^a$. Such eigenstates satisfy the properties
\cite{Pauli,Gen}
\be
&&v^a|iu\hb=iu^a|iu\hb,\;\;\;\vb -iu|=(|iu\hb)^{\dag}\nn\\
&&\vb iu'|iu\hb=\del^m(u'-u),\;\;\;\int d^mu|-iu\hb\vb -iu|=\int
d^mu|iu\hb\vb iu|=\bett
\e{29}
Notice that these relations requires $u^a$ to have symmetric ranges around
zero.
Inserting the last completeness relations in \r{23} and \r{24} we
obtain again using a normalization
like \r{26}:
 \be
&&_{+}\vb
ph|ph\hb_{+}=\left\{\begin{array}{c}\int d^mu\det{(R^{-1})^a}_b(u)
\vb\phi|e^{i\psi'_au^a}
|\phi\hb\\
\int d^mu \det{(L^{-1})^a}_b(u)\vb\phi|e^{i\psi''_au^a}|\phi\hb
\end{array}\right.
\nn\\
&&_{-}\vb
ph|ph\hb_{-}=(-1)^m\,_{+}\vb
ph|ph\hb_{+}
 \e{30}
where ${R^a}_b(u)={L^a}_b(-u)$ and where the last relation follows since
the ranges of
$u^a$ are symmetric around zero. These are the natural gauge theoretic
expressions:
$\det{(R^{-1})^a}_b(u)$ and $\det{(L^{-1})^a}_b(u)$ are the right and
left-invariant group measures.
$u^a$ canonical coordinates (which also are symmetric around zero) and
$e^{i\psi'_au^a}$ a finite
group element belonging to the identity component of the group. Equality
for positive and negative
$\al$'s requires that the normalization of the two matter vacua are
related by a factor $(-1)^m$ (see
section 4 of \cite{Propa}).

For unimodular gauge groups we have
\be
&&{U_{ab}}^b=0
\e{36}
which implies
\be
&&\psi''_a=\psi'_a=\psi_a
\e{361}
and
\be
&&\det{(L^{\dag-1})^a}_b(iv)\equiv
\det{(L^{-1})^a}_b(-iv)=\det{(L^{-1})^a}_b(iv)
\e{362}
In this case the two equivalent expressions on the right-hand side of
\r{27}, \r{28} and \r{30}
coincide. (In particular we have $\det R^{-1}=\det L^{-1}$ in \r{30}.)

 Exactly how one should choose to quantize the Lagrange
multipliers $v^a$ is not given beforehand. In fact it is not given how
to quantize any of the
involved variables beforehand. However, in \cite{Propa} the following
necessary condition for the
possible existence of the solution \r{1} was given (from the requirement
that \r{1} is an inner
product state): \be
&&\mbox{{\sl The  Lagrange multipliers $v^a$ must be quantized with
opposite metric states}}\nn\\
&&\mbox{{\sl to the
unphysical variables which the gauge generator $\psi_a$ eliminates.}}
\e{301}
For trivial models it was shown in \cite{Propa} that all such possible
choices lead to the same
result for a fixed choice of the quantization of the gauge invariant
physical matter variables. This
is not the case for more involved models where usually only {\em particular
choices make the formal
solutions exist}. In \cite{Propa} some simple, essentially abelian
examples were given. In the
following sections we shall treat  $SU(2)$ and $SL(2,R)$ gauge theories in
details.

 In order for the physical states to have positive norms
one must quantize the remaining matter variables with positive
metric states. This is  a necessary
condition. One may notice that this together with \r{301} require
$|\phi\hb$ in \r{30} to be
expandable in terms of positive metric states while this is not
the case in \r{27}-\r{28}. Maybe this
provide for the possibility to give a general proof that
the physical states \r{30} have positive
norms whenever they exist (which should   be the case at least
for compact gauge groups).

We end this section with a technical remark. In order to
calculate \r{27} and \r{30} explicitly one
has  to factorize the exponents $e^{\psi_av^a}$  and $e^{-\psi_av^a}$ in
\r{23} and \r{24}, \ie one has to determine new parameters $w_i$ in  relations
like
\be
&&e^{\psi_av^a}=e^{w^{\dag}_nA_n^{\dag}}\cdots
e^{w^{\dag}_2A_2^{\dag}}e^{w_1A_1}e^{w_2A_2}\cdots
e^{w_nA_n} \e{31}
where $A_i$ are linear expressions in $\psi_a$
($A_i\neq A_{i+1},A_{i-1}$), and     $w_i$
expressions in terms of the Lagrange multipliers $v^a$.   $A_1$ and $w_1$ must
be
hermitian since $e^{\psi_av^a}$ is hermitian. Notice that
$e^{-\psi_av^a}$ is the inverse operator to
$e^{\psi_av^a}$. We have therefore \be
&&e^{-\psi_av^a}=e^{-w_nA_n}\cdots e^{-w_2A_2}e^{-w_1A_1}
e^{-w^{\dag}_2A_2^{\dag}}\cdots
e^{-w^{\dag}_nA_n^{\dag}} \e{311}
Hence, the transformation $v^a\ra-v^a$ is accompanied by
$w_1\ra-w_1$ and $w_i\ra-w_i^{\dag}$ for
$i\geq2$.
Locally we shall always require that $v^a$ is
determined by $w_i$. This requires in turn that
 $n\geq(m+1)/2$
where $m$ is the number of Lagrange multipliers $v^a$. Now,
after a choice of quantization of the
original state space one will find that $w_i$ and $v^a$ are not
in a one-to-one correspondence if
the ranges are not restricted. These restrictions will depend on the
gauge group, the choice of
quantization and on the choice of factorization (the choice of
$A_i$ in \r{31}). Our results suggest
that the gauge group and the choice of quantization essentially
determines the appropriate
factorization. This is natural since the choice of factorization
is only a technical point which
cannot determine the quantum theory. A given gauge model determines,
of course, the factorization
and the possible choices of quantization. This procedure will be treated in
detail
for $SU(2)$ and $SL(2,R)$ gauge theories in the following sections.
Also a nonunimodular group will
be treated.

\setcounter{equation}{0}
 \section{SU(2) gauge theories}
In this section we give some general properties of an $SU(2)$ gauge theory.
Consider the case when
the gauge generators $\psi_a$ satisfy the Lie algebra (since the group metric
is
Euclidean we use only
lower indices)
\be
&[\psi_a, \psi_b]_{-}=i\varepsilon_{abc}\psi_c
\e{32}
where $\varepsilon_{abc}$ is the three dimensional antisymmetric symbol
($\varepsilon_{123}=1$). The
BRST charge operator is then \be
&&Q=\psi_a\eta_a-\frac{1}{2}i\varepsilon_{abc}\pet_a
\eta_b\eta_c + \bapet_a\pi_a
\e{33}
With the gauge fixing $\rho=\pet_av_a$ we find
\be
&&[\rho, Q]_+=\psi_av_a+\halv
i\varepsilon_{abc}(\pet_c\eta_b-\eta_b\pet_c)v_a+i\pet_a\bapet_a
\e{34}
For compact semisimple gauge groups, which are unimodular, the norms of
the physical states are
determined by the expressions
 \be
&&_{+}\vb ph|ph\hb_{+}=\vb\Phi|e^{[\rho,
Q]_+}|\Phi\hb=\vb\phi|\,_{\pi}\vb0|e^{\psi_av_a}
\det{(L^{-1})}_{ab}(iv)|0\hb_{\pi}|\phi\hb\nn\\
&&_{-}\vb ph|ph\hb_{-}=\vb\Phi|e^{-[\rho,
Q]_+}|\Phi\hb=(-1)^m\vb\phi|\,_{\pi}\vb0|e^{-\psi_av_a}
\det{(L^{-1})}_{ab}(iv)|0\hb_{\pi}|\phi\hb
 \e{35}
from \r{23}-\r{24} and \r{36}-\r{362}.
For $SU(2)$ the left invariant vielbein $(L^{-1})_{ab}$ has the explicit form
($v=\sqrt{v_1^2+v_2^2+v_3^2}$)
 \be
&&(L^{-1})_{ab}(iv)=\delta_{ab}\frac{\sinh v}{v}-\frac{v_av_b}{v^2}
(\frac{\sinh v}{v}-1)
-i\varepsilon_{abc}v_c\frac{\cosh v-1}{v^2}
\e{37}
from which we find the measure operator
\be
&&\det(L^{-1})_{ab}(iv)=\frac{2(\cosh v-1)}{v^2}
\e{38}
which satisfies the unimodularity properties \r{36} and \r{362}. It is
natural to restrict the
possible spectra of $v_a$ to those which make the measure operator \r{38} real.
Here this requires
that $v_a$ for each value of $a$ should have either real or imaginary
eigenvalues.
For real spectra
of all three Lagrange multipliers $v_a$ we find
 \be
&&_{+}\vb
ph|ph\hb_{+}=\int d^3v\frac{2(\cosh v-1)}{v^2}\vb\phi|e^{\psi_av_a}
|\phi\hb\nn\\
&&_{-}\vb
ph|ph\hb_{-}=-\int d^3v\frac{2(\cosh v-1)}{v^2}\vb\phi|e^{-\psi_av_a}
|\phi\hb
 \e{39}
and for imaginary spectra ($v_a\ra iu_a$, $u=\sqrt{u_1^2+u_2^2+u_3^2}$)
 \be
&&_{+}\vb
ph|ph\hb_{+}=\int d^3u\frac{2(1-\cos u)}{u^2}\vb\phi|e^{i\psi_au_a}
|\phi\hb\nn\\
&&_{-}\vb
ph|ph\hb_{-}=-_{+}\vb
ph|ph\hb_{+}
 \e{40}
These are the two natural choices for an SU(2) gauge theory since we
 have to restrict the spectra of
$v^a$ so that $v=\sqrt{v_1^2+v_2^2+v_3^2}$ is either real or imaginary.
(The transition between
these two possibilities is double valued, $v\ra\pm iu$.) Furthermore, since
the last measure is zero
for $u=2\pi n,\;n=\pm1,\pm2,...$ it seems in this case also  natural to
restrict the ranges of
$u_a$ within the limits $0\leq u<2\pi$.

In order to calculate the matrix elements $\vb\phi|e^{\pm\psi_av_a}
|\phi\hb$ in \r{39}-\r{40} one
needs in general to factorize the exponential operator \r{35} like \r{31}.
We may \eg
choose the Euler angle like decomposition
\be
&&e^{\psi_av_a}=e^{\al\psi_3}e^{\beta\psi_2}e^{\ga\psi_3}
\e{41}
where $\al, \beta$ and $\ga$ are expressions in terms of $v_a$.
The right-hand side is manifestly
hermitian if
\be
&&\beta^{\dag}=\beta,\;\;\;\ga=\al^{\dag}
\e{42}
which we require. Eq.\r{41} is easily solved algebraically. We find
\be
\frac{v^1}{v}\sinh\frac{v}{2}=-i\sinh {\frac{\beta}{2}}
\sinh{\frac{\alpha-\gamma}{2}},\;\;\;
\frac{v^2}{v}\sinh\frac{v}{2}=\sinh {\frac{\beta}{2}}
\cosh{\frac{\alpha-\gamma}{2}}\nn\\
\frac{v^3}{v}\sinh\frac{v}{2}=\cosh {\frac{\beta}{2}}
\sinh{\frac{\alpha+\gamma}{2}},\;\;\;
\cosh\frac{v}{2}=\cosh{\frac{\beta}{2}}\cosh{\frac{\alpha+\gamma}{2}}.
\e{461}
All these relations are consistent with hermitian $v_a$ and \r{42}.

For the case \r{39} we find then with $\al=\xi+i\zeta$ and $\xi$, $\zeta$ real
 \be
&&_{+}\vb
ph|ph\hb_{+}=\halv\int d\xi d\beta d\zeta
\sinh\beta\vb\phi|e^{(\xi+i\zeta)\psi_3}e^{\beta\psi_2}
e^{(\xi-i\zeta)\psi_3}|\phi\hb\nn\\
&&_{-}\vb
ph|ph\hb_{-}=-_{+}\vb
ph|ph\hb_{+}
 \e{47}
where $\beta\geq0$ and $-\infty <\xi,\zeta <\infty$ for infinite ranges
of the Lagrange
multipliers $v_a$.

For the case \r{40} we find on the other hand the standard Euler angle
decomposition
($\al, \beta,
\ga$ real)
 \be
&&_{+}\vb
ph|ph\hb_{+}=\int d\al d\beta d\ga
\sin\beta\vb\phi|e^{i\al\psi_3}e^{i\beta\psi_2} e^{i\ga\psi_3}|\phi\hb\nn\\
&&_{-}\vb
ph|ph\hb_{-}=-_{+}\vb
ph|ph\hb_{+}
 \e{48}
where it is natural to restrict the ranges to the group manifold.
$0\leq\beta\leq\pi$,
$-\pi\leq\al,\ga\leq\pi$ (SO(3)) or $-2\pi\leq\al\pm\ga\leq2\pi$ (SU(2))
which corresponds to $0\leq
u\leq\pi$ in \r{40}.

\setcounter{equation}{0}
\section{SL(2,R) gauge theories}
In this section we consider some general properties of an $SL(2,R)$ gauge
theory.
We consider the
case when  the gauge generators $\psi_a$ satisfy the algebra
\be
&[\psi_a, \psi_b]_{-}=i{\varepsilon_{ab}}^c\psi_c
\e{50}
where we make use of  the pseudo-euclidean metric $\eta_{ab}$ (diag
$\eta_{ab}=(-1,1,1)$). The BRST
charge operator is
\be
&&Q=\psi_a\eta^a-\frac{1}{2}i{\varepsilon_{bc}}^a\pet_a
\eta^b\eta^c + \bapet_a\pi^a
\e{51}
With the gauge fixing $\rho=\pet_av^a$ we find
\be
&&[\rho, Q]_+=\psi_av^a+ \halv
i{\varepsilon_{ab}}^c(\pet_c\eta^b-\eta^b\pet_c)v^a+i\pet_a\bapet^a
\e{52}
and from the general formulas \r{15},\r{22},\r{23}, and \r{24} we get
 \be
&&_{+}\vb ph|ph\hb_{+}=\vb\Phi|e^{[\rho,
Q]_+}|\Phi\hb=\vb\phi|\,_{\pi}\vb0|e^{\psi_av^a}
\det{(L^{-1})^a}_b(iv)|0\hb_{\pi}|\phi\hb\nn\\
&&_{-}\vb ph|ph\hb_{-}=\vb\Phi|e^{-[\rho,
Q]_+}|\Phi\hb=\vb\phi|\,_{\pi}\vb0|e^{-\psi_av^a}
\det{(L^{-1})^a}_b(iv)|0\hb_{\pi}|\phi\hb
 \e{53}
For $SL(2,R)$ the left invariant vielbein ${(L^{-1})^a}_b(iv)$ has the explicit
form\\
($v=\sqrt{-(v^1)^2+(v^2)^2+(v^3)^2}$)
\be
&&{(L^{-1})^a}_b(iv)=\delta^a_{b}\frac{\sinh v}{v}+\frac{v^av_b}{v^2}
(1-\frac{\sin v}{v})
-i{\varepsilon^a}_{bc}v^c\frac{\cos v-1}{v^2}
\e{54}
from which we find the measure operator
\be
&&\det{(L^{-1})^a}_{b}(iv)=\frac{2(1-\cos v)}{v^2}
\e{55}
which satisfies the properties \r{36} and \r{362}. Even in this case
the eigenvalue of the measure
operator is real if $v^a$ for each value of $a$ has either real or
imaginary eigenvalues. For real
spectra of the Lagrange multipliers $v^a$ we find
($v\equiv\sqrt{-(v^1)^2+(v^2)^2+(v^3)^2}$ )
 \be
&&_{+}\vb
ph|ph\hb_{+}=\int d^3v\frac{2(1-\cos v)}{v^2}\vb\phi|e^{\psi_av^a}
|\phi\hb\nn\\
&&_{-}\vb
ph|ph\hb_{-}=-\int d^3v\frac{2(1-\cos v)}{v^2}\vb\phi|e^{-\psi_av^a}
|\phi\hb
 \e{56}
and for imaginary spectra ($v^a\ra iu^a$, $u=\sqrt{-(u^1)^2+(u^2)^2+(u^3)^2}$)
 \be
&&_{+}\vb
ph|ph\hb_{+}=\int d^3u\frac{2(\cosh u-1)}{u^2}\vb\phi|e^{i\psi_au^a}
|\phi\hb\nn\\
&&_{-}\vb
ph|ph\hb_{-}=-_{+}\vb
ph|ph\hb_{+}
 \e{57}
Now as in the SU(2) case we require the eigenvalues of
$v=\sqrt{-(v^1)^2+(v^2)^2+(v^3)^2}$ to be
either real or imaginary since the transition $v\ra\pm iu$ is double valued.
This is accomplished in
\r{56} by restrictions of the type $v^2\pm v^1\geq0$. It is also obtained
if the quantization
is chosen so that $v^1$ has imaginary eigenvalues ($v^1=iu^1$) and $v^2$, $v^3$
real ones
($v=\sqrt{(u^1)^2+(v^2)^2+(v^3)^2}$). In this case
it is also natural to restrict the ranges further so that $v$ lies in the range
$0\leq v<2\pi$ since
the measure is zero for $u=2\pi n$ ($n=\pm1,\pm2,...$). However, there are no
allowed restrictions on
$u^a$ which make $u$ real in \r{57} since the ranges of $u^a$ must be symmetric
around zero. Thus, in
distinction to the SU(2) case one may not choose the natural unitary
quantization \r{57} in an
SL(2,R) gauge theory! However, a real $u$ is obtained for the following
choices of the
quantization:\ben \item $v^1$ has real eigenvalues and $v^2$, $v^3$ imaginary
ones\\
($u=\sqrt{(v^1)^2+(u^2)^2+(u^3)^2}$)
\item $v^1$, $v^2$ have real eigenvalues and $v^3$ imaginary ones together
with the restriction
$v^1\pm v^2\geq0$
\item $v^1$, $v^3$ have real eigenvalues and $v^2$ imaginary ones together
with the restriction
$v^1\pm v^3\geq0$
\een
together with combinations of cases 2) and 3). There are no further
restrictions in these cases.

We may now proceed  and analyze the above cases further by a
factorization of the exponential
$e^{\psi_av_a}$ in \r{53}. For the case when $v^a$ have real
eigenvalues and when $v$ is real it
is natural to consider the factorization (case II in Appendix B)
\be
&&e^{\psi_av_a}=e^{\al\psi_3}e^{\beta\psi_2}e^{\ga\psi_3}
\e{64}
Real $v^a$ requires real $\beta$ and $\al=\xi+i\zeta$, $\ga=\xi-i\zeta$
where $\xi$ and $\zeta$
are real with the ranges $-\pi\leq\xi\leq\pi$, $-\infty <\zeta <\infty$.
The restriction
$v^2\pm v^1\geq0$ requires $0\leq\beta\leq\pi$. Eq.\r{56} reduces to
 \be
&&_{+}\vb
ph|ph\hb_{+}=\int d\xi d\zeta d\beta
\sin\beta\vb\phi|e^{(\xi+i\zeta)\psi_3}e^{\beta\psi_2}e^{(\xi-i\zeta)\psi_3}
|\phi\hb
\e{65}

In the case when $v^1$ has imaginary eigenvalues and $v^2$, $v^3$ real ones,
one may also make use
of the factorization \r{64} but now with $\al, \beta, \ga$ real and with
the ranges
$-\pi\leq\al\pm\ga\leq\pi,\; 0\leq\beta\leq\pi$ which actually requires
the additional restriction
$v^2>0$. Eq.\r{56} reduces then to
 \be
&&_{+}\vb
ph|ph\hb_{+}=\int d\al  d\beta d\ga
\sin\beta\vb\phi|e^{\al\psi_3}e^{\beta\psi_2}e^{\ga\psi_3} |\phi\hb
\e{66}
Alternatively one may make use of the factorization (case III in Appendix B)
\be
&&e^{\psi_av_a}=e^{\al\psi_1}e^{\beta\psi_3}e^{\ga\psi_1}
\e{67}
which requires $v^3>0$. Here the eigenvalues of $\al$ and $\ga$ must be
imaginary and eq.\r{56}
reduces to ($\al, \ga \ra i\al, i\ga$)
 \be
&&_{+}\vb
ph|ph\hb_{+}=\int d\al  d\beta d\ga
\sin\beta\vb\phi|e^{i\al\psi_1}e^{\beta\psi_3}e^{i\ga\psi_1} |\phi\hb
\e{68}
with the same ranges as in \r{66}.

In the case when $v^1$ is real and $v^2, v^3$ imaginary and $u$ real
(case 1 above) we may use the
factorization (case I in Appendix B)
\be
&&e^{\psi_av_a}=e^{\al\psi_3}e^{\beta\psi_1}e^{\ga\psi_3}
\e{69}
Here $\beta$ is real and should be positive or negative. For $\beta>0$
we must impose the restriction
$v^1>0$. For $\al=\xi+i\zeta$ and $\ga=-\xi+i\zeta$ with $\xi,\zeta$ real
the ranges are
$-\pi/2<\xi<\pi/2,\;-\infty<\zeta<\infty$. Eq.\r{57} reduces to
 \be
&&_{+}\vb
ph|ph\hb_{+}=\int d\xi d\zeta d\beta \sinh\beta
\vb\phi|e^{(\xi+i\zeta)\psi_3}e^{\beta\psi_1}e^{(-\xi+i\zeta)\psi_3} |\phi\hb
\e{70}

For the case when $v^1, v^2$ are real and $v^3$ imaginary with $u$ real
 and $v^1\pm v^2>0$ or $v^1\pm
v^2<0$ (case 2 above) there are several possible factorizations. We may use
\r{69}
with $\beta$ real and $\al$ and $\ga$ imaginary. The ranges of
$\al$ and $\ga$ are then infinite
while $\beta>0$ for $v^1\pm v^2>0$. Eq.\r{57} reduces to
($\al, \ga \ra i\al, i\ga$)
 \be
&&_{+}\vb
ph|ph\hb_{+}=\int d\al  d\beta d\ga \sinh\beta
\vb\phi|e^{i\al\psi_3}e^{\beta\psi_1}e^{i\ga\psi_3} |\phi\hb
\e{71}
In this case we may also make use of the factorizations IV and V in Appendix B,
\ie
\be
&&e^{\psi_av_a}=e^{\al \phi_1}e^{\beta \phi_2}e^{\ga \phi_1},\;\;\;
e^{\psi_av_a}=e^{\al \phi_2}e^{\beta \phi_1}e^{\ga \phi_2}
\e{72}
where
\be
&&\phi_1=\frac1{\sqrt2}(\psi_1+\psi_2),\;\;\;
\phi_2=\frac1{\sqrt2}(\psi_1-\psi_2)
\e{73}
$\al, \beta$ and $\ga$ are real here and  are \eg positive for $v^1\pm v^2>0$.
Eq. \r{57} reduces to
 \be
&&_{+}\vb
ph|ph\hb_{+}=\int d\al  d\beta d\ga \beta
\vb\phi|e^{\al\phi_1}e^{\beta\phi_2}e^{\ga\phi_1} |\phi\hb
\e{731}
\be
&&_{+}\vb
ph|ph\hb_{+}=\int d\al  d\beta d\ga \beta
\vb\phi|e^{\al\phi_2}e^{\beta\phi_1}e^{\ga\phi_2} |\phi\hb
\e{74}

At this general level one cannot proceed further except to discuss more
factorizations. Now it
depends on the properties of the specific model whether or not any
of the above cases lead to finite
(positive) physical norms.

\setcounter{equation}{0}
\section{A specific SL(2,R) model}
Consider a finite dimensional model with dimension $n$ spanned by the
coordinates $y^A$ and momenta
$p_A$ ($A=1,...,n$). Let the indices be raised and lowered by the constant
symmetric metric matrices
$\eta^{AB}$ and $\eta_{AB}$ respectively ($\eta^{AB}\eta_{BC}=\del^A_C$).
On this phase space one may
naturally introduce an SL(2,R) gauge symmetry which is invariant under
the global symmetry implied by
the metric $\eta^{AB}$ through the following constraints
\be
&&p^2\equiv p_Ap_B\eta^{AB}=0,\;\;\;y^2\equiv y^Ay^B\eta_{AB}=0,
\;\;\;p\cdot y\equiv p_Ay^A=0
\e{75}
Such a model is described by the phase space Lagrangian
\be
&&L=p_A\dot{y}^A-\la_1p^2-\la_2y^2-\la_3p\cdot y
\e{76}
where $\la_a$ are Lagrange multipliers. A configuration space Lagrangian is
obtained by  an
elimination of $p_A$ which is possible if $\la_1\neq0$ which we assume. In
\cite{MA,MN} such a
model with an O(4,2) metric was proposed and analyzed. It describes both
a massless and a massive
free relativistic particle depending on the choice of gauge.

Imposing the canonical commutation relations
\be
&&[y^A, p_B]_-=i\del^A_B
\e{77}
one easily finds that the normalized constraint operators
\be
&&\phi_1=\frac{1}{2\sqrt2}p^2,\;\;\;\phi_2=\frac{1}{2\sqrt2}y^2,\;\;\;
\phi_3=\frac{1}{4}(p\cdot
y+y\cdot p)
\e{78}
satisfy the commutator algebra
\be
&&[\phi_1, \phi_2]_-=-i\phi_3,\;\;\;[\phi_2, \phi_3]_-=i\phi_2,\;\;\;
[\phi_3, \phi_1]_-=i\phi_1
\e{79}
This is an SL(2,R) algebra. To see this one may consider the following
redefined constraint operators
\be
&&\psi_1=\frac{1}{\sqrt2}(\phi_1+\phi_2)=\frac{1}{4}(p^2+y^2),
\;\;\;\psi_2=\frac{1}{\sqrt2}(\phi_1-\phi_2)=\frac{1}{4}(p^2-y^2),\nn\\
&&\psi_3=\phi_3=\frac{1}{4}(p\cdot
y+y\cdot p)
\e{80}
$\psi_a$ satisy then the standard SL(2,R) algebra \r{50}.

Now one may notice that $\phi_i$ in \r{78} have exactly the same relation
to $\psi_a$ as
$\phi_{1,2}$ in \r{73}. Hence, the factorizations \r{72} are very natural
choices here. The
expressions \r{731} and \r{74} may be integrated (for $n\geq 5$) after
insertion of the completeness
relations \be
&&\int d^n y |y\hb\vb y|=\bett,\;\;\;\int d^n p |p\hb\vb p|=\bett
\e{801}
for real eigenvalues and with the modifications implied by \r{29} for
imaginary eigenvalues.
We find
\be
&&_{\pm}\vb
ph|ph\hb_{\pm}=\int d^n y d^n y'\frac{\phi(y)^*}{y^2}
\frac{1}{(y-y')^{2(\frac{n}{2}-2)}}
\frac{\phi(y')}{y'^2}\nn\\
&&_{\pm}\vb
ph|ph\hb_{\pm}=\int d^n p d^n p'\frac{\phi(p)^*}{p^2}
\frac{1}{(p-p')^{2(\frac{n}{2}-2)}}
\frac{\phi(p')}{p'^2}
\e{802}
for appropriate choices of the signs of $\al,\beta$ and $\ga$
(which is opposite for $|ph\hb_+$ and
$|ph\hb_-$) and provided $y^2>0$, ${y'}^2>0$, $(y-y')^2>0$ and
$p^2>0$, ${p'}^2>0$, $(p-p')^2>0$
respectively. Since two of the Lagrange multipliers have real
eigenvalues, the quantization rule
\r{301} requires two of the coordinate and momentum components
to be quantized with imaginary
eigenvalues. This together with the requirement $y^2>0$ and
$p^2>0$ etc imply that \r{802} is only
valid for models with a global $O(n-2,2)$ invariance where
$n\geq 5$. Also the factorization \r{69}
is possible for the above cases and should yield the result
\r{802} as well. However, \r{71} is more
complicated than \r{731}-\r{74} and we have not worked it out
explicitly. We notice though that the
$\beta$ integration is only finite for $p^2+y^2>0$ and for
dimensions $n\geq 5$ (the vacuum energy
for $p^2+y^2$ is $n/2$ and yields $e^{\beta\frac{n}{4}}$
which together with the factor $e^{-\beta}$
from $\sinh\beta$ must lead to $e^{\beta\del}$, $\del>0$).

The physical norms \r{802} are actually positive. This is
easily seen by means of a Fourier
transformation. We find
\be
&&_{\pm}\vb
ph|ph\hb_{\pm}=\int d^n p
{{\tilde\phi}(p)^*}\frac{1}{(p^{2})^2}{{\tilde\phi}(p)}\nn\\
&&_{\pm}\vb
ph|ph\hb_{\pm}=\int d^n y {{\tilde\phi}(y)^*}\frac{1}{(y^2)^{2}}
{{\tilde\phi}(y)}
\e{8020}
where ${\tilde\phi}(p)$ and ${\tilde\phi}(y)$ are Fourier
transforms of $\phi(y)/y^2$ and
$\phi(p)/p^2$ respectively.

In six dimensions ($n=6$) \r{802} reduces to
\be
&&_{\pm}\vb
ph|ph\hb_{\pm}=\int d^6 y d^6 y'\frac{\phi(y)^*}{y^2}\frac{1}{(y-y')^{2}}
\frac{\phi(y')}{y'^2}\nn\\
&&_{\pm}\vb
ph|ph\hb_{\pm}=\int d^6 p d^6 p'\frac{\phi(p)^*}{p^2}\frac{1}{(p-p')^{2}}
\frac{\phi(p')}{p'^2}
\e{803}
which from the above argument is valid for a model with a global O(4,2)
invariance \ie exactly the
invariance of the model in \cite{MA,MN}. In  this model we may look more
precisely into the
connection between the quantization of Lagrange multipliers and matter
coordinates as prescribed by
the quantization rule \r{301}. In \cite{MA,MN} the components $A$
were chosen to run over the values
$A=0,1,2,3,5,6$ where $y^0$ is (related) to the time component. In
the reduction to a massive or
massless free relativistic particle $p^2,\;p\cdot y,\;y^2$ eliminate
the coordinates $y^0,\;y^5$ and
$y^6$ respectively. Since $v^1,\;v^2$ are real and $v^3$ imaginary
we should according to the
rule \r{301} choose $y^0,\;y^6$ imaginary which is exactly what is
required to have $y^2>0$ and
$p^2>0$ in \r{803} since $\eta_{00}=\eta_{66}=-1$.  Since time is
imaginary \r{803} represents a
euclidean expression for the free relativistic particle both in the
massive and in the massless case.

\setcounter{equation}{0}
\section{An example of a nonunimodular gauge group; the scaling group}
A simple gauge group which is not unimodular is generated
by $\psi_1$ and $\psi_2$ satisfying the
Lie algebra
\be
&&[\psi_1, \psi_2]=i\psi_2
\e{81}
Since $\psi_1$ generates scaling of $\psi_2$ we call this group
the scaling group. The BRST charge
operator \r{3} is in this case
\be
&&Q=\psi_a\eta^a-i\pet_2\eta^1\eta^2-\frac{i}{2}\eta^1+\pi_a\bapet^a
\e{82}
The gauge fixing operator $\rho=\eta^a\pet_a$ leads then to the
expressions \r{15} and \r{18} where
\be
&&{(L^{-1})^a}_b(iv)=\del^a_1\del^1_b+\del^a_2\del^2_b\frac{i}{v^1}(1-e^{iv^1})+\nn\\
&&+\del^a_1\del^2_b\frac{v^2}{v^1}(\frac{i}{v^1}(e^{iv^1}-1)+1)
\e{83}
The scaling group is not unimodular since
\be
&&\det{(L^{\dag-1})^a}_b\neq\det{(L^{-1})^a}_b,\;\;\;{U_{1a}}^a=1
\e{84}
which violates \r{36}. As a consequence we obtain the two formally
different expressions in \r{23}
and \r{24} for $_{\pm}\vb ph|ph\hb_{\pm}$. However, inserting the
explicit expression \r{83} we find
the unique answer
\be
&&_{\pm}\vb ph|ph\hb_{\pm}=\vb\phi|\,_{\pi}\vb0|e^{\pm\psi_av^a}
\frac{2}{v^1}\sin\frac{v^1}{2}|0\hb_{\pi}|\phi\hb
\e{85}
Thus, the difference in $\det{(L^{\dag-1})^a}_b$ and $\det{(L^{-1})^a}_b$ is
 compensated by the factors caused by ${U_{ab}}^b\neq0$. For real
spectra of $v^a$ we have in
particular
\be
&&_{\pm}\vb ph|ph\hb_{\pm}=\int dv^1 dv^2 \frac{2}{v^1}
\sin\frac{v^1}{2}\vb\phi|e^{\pm\psi_av^a}
|\phi\hb
\e{86}
and for imaginary spectra ($v^a=iu^a$)
\be
&&_{\pm}\vb ph|ph\hb_{\pm}=\int du^1 du^2 \frac{2}{u^1}
\sinh\frac{u^1}{2}\vb\phi|e^{\pm i\psi_au^a}
|\phi\hb
\e{87}
We may also have a mixture, \ie $v^1$ real and $v^2$ imaginary or vice versa.
For $v^1$ real the
measure is zero for $v^1=2\pi n$ where $n$ is a nonzero integer. Hence,
in this case it is natural
to restrict $v^1$ to \eg $-2\pi<v^1<2\pi$.

To proceed and to make still simpler expressions we need to factorize
$e^{\pm\psi_av^a}$. It is
straight-forward to derive
\be
&&e^{\pm\psi_av^a}=e^{\pm\frac{v^1}{2}\psi_1}e^{\pm\beta\psi_2}
e^{\pm\frac{v^1}{2}\psi_1}
\e{88}
where
\be
&&\beta=\left(\frac{2}{v^1}\sin\frac{v^1}{2}\right)v^2
\e{89}
which is real or imaginary for $v^2$ real or imaginary. Hence,
depending on the spectra of the
Lagrange multipliers $v^a$ we arrive at the following formulas
for $_{\pm}\vb ph|ph\hb_{\pm}$: For
real $v^a$
 \be
&&_{\pm}\vb ph|ph\hb_{\pm}=\int dv^1 d\beta
\vb\phi|e^{\pm\frac{v^1}{2}\psi_1}e^{\pm\beta\psi_2}e^{\pm
\frac{v^1}{2}\psi_1}|\phi\hb
\e{90}
and for imaginary $v^a$ we have ($v^1\ra iu^1$, $\beta\ra i\beta$)
 \be
&&_{\pm}\vb ph|ph\hb_{\pm}=\int du^1 d\beta
\vb\phi|e^{\pm i\frac{u^1}{2}\psi_1}e^{\pm i\beta\psi_2}e^{\pm i
\frac{u^1}{2}\psi_1}|\phi\hb
\e{91}
For the mixtures we have the obvious combinations.

\setcounter{equation}{0}
\section{Explicit realizations of the scaling group}
Consider a matter state space spanned by the hermitian coordinate
and momentum operators $y^A$ and
$p_A$ ($A=1,\ldots,n$) satisfying the commutation relations \r{77}.
As in section 5 we let the
indices be raised and lowered by the real constant symmetric metric
$\eta^{AB},\;\eta_{AB}$ ($\eta^{AB}\eta_{BC}=\del^A_C$). On this phase
space the hermitian gauge
generators $\psi_1$ and $\psi_2$ satisfying the
Lie algebra \r{81} may be realized by the expressions
\be
&&\psi_1=\frac{1}{4}(p\cdot y+y\cdot p),\;\;\;\psi_2=\al p^2
\e{92}
where $\al$ is a real constant different from zero. In \cite{MAN} a
five dimensional model of this
kind was proposed. It has an O(4,1) metric and was shown to describe
either a free massive
relativistic particle or a massless one in a de Sitter space depending
on the choice of gauge.

In terms of this explicit realization we may now further simplify
the general formulas
\r{90}-\r{91}. Due to the form of $\psi_2$ in \r{92} it is natural to
write these expressions in
terms of wave functions in momentum space by means of the completeness relation
\be
&&\int d^np |p\hb\vb p|=\bett
\e{93},
for real $p_A$ and with the modification implied by \r{29} for imaginary ones.
Since
\be
&&\vb p|e^{\pm\frac{v^1}{2}\psi_1}|\phi\hb=\vb e^{\pm i\frac{v^1}{4}}p|\phi\hb
\e{94}
only imaginary spectra of $v^1$ leads to wave functions with real arguments.
For imaginary $v^1$ and
$v^2$ ($v^a\ra iu^a$) we find from \r{91}
\be
&&_{\pm}\vb ph|ph\hb_{\pm}=\int du^1 d\beta d^np\,
\phi^*(e^{\pm \frac{u^1}{4}}p) e^{\pm i\beta\al p^2}\phi(e^{\mp
\frac{u^1}{4}}p)=\nn\\
&&=\frac{4}{|\al|}\int d\ga d^np\, \phi^*(e^{\pm \ga}p) \del(p^2)\phi(e^{\mp
\ga}p)
\e{95}
and for  real $v^2$ restricted to positive or negative eigenvalues
\be
&&_{\pm}\vb ph|ph\hb_{\pm}=\frac{4}{|\al|}\int d\ga \frac{d^np}{p^2}
\phi^*(e^{\pm \ga}p)
\phi(e^{\mp \ga}p)
\e{96}
This construction requires $p^2>0$ and since only one momentum
component is quantized with imaginary
eigenvalues according to the general quantization rule, the expression
\r{96} is only valid for an
$O(n-1,1)$ metric $\eta_{AB}$ and in particular for the $O(4,1)$ model
considered in \cite{MAN}.
Although all momentum components are to be quantized with real eigenvalues
in \r{95} only an
$O(n-1,1)$ metric makes the scalar product \r{95} nonzero and well defined.
Thus, only for an
$O(n-1,1)$ metric $\eta_{AB}$ do we find well defined physical scalar products.
We claim that the
norms \r{95} and \r{96} are positive when they exist.

\setcounter{equation}{0}
\section{Conclusions}
In this paper we have dealt with both  general and specific
properties of gauge theories with
finite number of degrees of freedom and when quantized on inner
product spaces by means of a BRST
procedure. For Lie group theories we  derived general geometrical
expressions for the norms of
the physical states found in \cite{Simple}. We obtained expressions
involving only
matter states, gauge generators and vielbeins on the group manifold,
\r{27}-\r{28}. When the Lagrange
multipliers were quantized with indefinite metric states we obtained
the most natural geometrical
expressions \r{30} which involve the group measures and which we expect
to be useful for compact Lie
groups like $SU(2)$. However, for noncompact Lie groups like $SL(2,R)$
these natural expressions are
inappropriate. In this case some Lagrange multipliers must be quantized
with positive metric states.
We have performed a detailed analysis of  $SL(2,R)$ theories in order to
determine the
possible quantization choices of the Lagrange multipliers. These formulas
were then applied to a
simple class of $SL(2,R)$ theories. In this way we found physical states of the
BRST
quantization for these models  which  have positive norms. It
would be nice if our general analysis of
the $SL(2,R)$ case could be endowed with a natural geometrical
setting which in turn could serve as a
guide to more complex noncompact gauge groups.

Since our general formulas are also valid for gauge groups which
are not unimodular we analysed the
simple scaling group which involves two generators and applied the
resulting formulas to a simple
class of models with this invariance.  Physical states
with positive norms were then obtained.

For the cases in which the general formulas \r{27}-\r{28} are not
applicable we expect that the
procedure of ref.\cite{Gauge} should work instead. This procedure
involves a Dirac quantization. If
neither \r{27}-\r{28} nor ref.\cite{Gauge} nor any combination of
the two procedures is applicable we
believe that a BRST quantization is not possible at all.\\
\vspace{2cm}\\
{\bf Acknowledgement}

It is a pleasure to thank Arne Kihberg for helpful comments.

\setcounter{section}{1}
\setcounter{equation}{0}
\renewcommand{\thesection}{\Alph{section}}
\newpage
\noindent
{\Large{\bf{Appendix A}}}
 \vspace{5mm}\\
{\bf Derivation of the general expressions \r{15} and \r{18} in section 2.}

{}From \r{14} we have
 \be
   &&e^{[\rho, Q]_+}=e^{\psi'_av^a}e^{a+b}=e^{\psi''_av^a}e^{a+c}
\e{a1}
where
\be
&&\psi'_a\equiv\psi_a-\halv i{U_{ab}}^b,\;\;\;\psi''_a\equiv\psi_a+\halv
i{U_{ab}}^b=(\psi'_a)^{\dag},\nn\\
&&a\equiv -i\pet_a\bapet^a,\;\;\;b\equiv i{U_{ab}}^c\pet_c\eta^b,\;\;\;c\equiv
-i{U_{ab}}^c\eta^b\pet_c=b^{\dag}
 \e{a2}
Define now
\be
&&F(\la)\equiv e^{\la(a+b)}e^{-\la b},\;\;\;G(\la)\equiv e^{-\la c}e^{\la(a+c)}
\e{a3}
Then we have
\be
&&F(1)|\Phi\hb=e^{a+b}|\Phi\hb,\;\;\;
\vb\Phi|G(1) =\vb\Phi|e^{a+c}
\e{a4}
due to the property \r{10}. $F(\la)$ and $G(\la)$ are easily calculated. We
have \eg
\be
&&F'(\la)=F(\la)A(\la)=B(\la)F(\la)
\e{a5}
where
\be
&&A(\la)\equiv e^{\la b}ae^{-\la b},\;\;\;B(\la)\equiv e^{\la(a+
b)}ae^{-\la(a+b)}
\e{a6}
One may easily show that
\be
&&A(\la)=B(\la)=e^{\la[\rho,Q]}ae^{-\la[\rho,Q]}
\e{a7}
Hence
\be
&&F(\la)=e^{\int_0^{\la}d\la'A(\la')}
\e{a8}
$A(\la)$ is explicitly (see also \cite{Simple})
\be
&&A(\la)=-i{A^a}_b(i\la v)\pet_a\bapet^b
\e{a9}
where ${A^a}_b(i\la v^a)$ is the adjoint matrix representation of the
gauge group with the imaginary group
coordinates $i\la v^a$. Now we have the formula
\be
&&{(L^{-1})^a}_b(iv)=\int_0^1d\la {A^a}_b(i\la v)
\e{a10}
where ${L^a}_b(iv)$ is the left invariant vielbein on the group
manifold with imaginary coordinates
$iv^a$. Hence
\be
&&F(1)=e^{-i{(L^{-1})^a}_b(iv)\pet_a\bapet^b}
\e{a11}
Similarly one may show that
\be
&&G(1)=e^{-i{(L^{\dag-1})^a}_b(iv)\pet_a\bapet^b}=(F(1))^{\dag}
\e{a12}
where
\be
&&{L^{\dag a}}_b(iv)={L^{ a}}_b(-iv)
\e{a13}
is the right invariant vielbein on the group manifold. Formulas \r{15} follow
now.

For $e^{-[\rho, Q]_+}$ we have
\be
   &&e^{-[\rho, Q]_+}=e^{-\psi'_av^a}e^{-a-b}=e^{-\psi''_av^a}e^{-a-c}
\e{a14}
which implies
\be
&&F(-1)|\Phi\hb=e^{-a-b}|\Phi\hb,\;\;\;\vb\Phi|G(-1)=\vb\Phi|e^{-a-c}
\e{a15}
where
\be
&&G(-1)=e^{-\int_0^1d\la
A(\la)}=e^{i{(L^{-1})^a}_b(iv)\pet_a\bapet^b},\nn\\
&&F(-1)=e^{i{(L^{\dag-1})^a}_b(iv)\pet_a\bapet^b}=(G(-1))^{\dag}
\e{a16}
Formulas \r{18} follow now.\\
\vspace{1cm}

\setcounter{section}{2}
\setcounter{equation}{0}
\noindent
{\Large{\bf{Appendix B}}}
 \vspace{5mm}\\
{\bf Some factorizations for SL(2,R).}

Let $\psi_a$ be the gauge generators for SL(2,R) satisfying the algebra \r{50}.
The
following factorizations of  $e^{\psi_av_a}$ in \r{53} are considered in
section 4:\\
{\bf Case I}:
\be
&&e^{\psi_av_a}=e^{\al\psi_3}e^{\beta\psi_1}e^{\ga\psi_3}
\e{b1}
where $\al, \beta$ and $\ga$ are expressions in terms of $v_a$.
The right-hand side is manifestly
hermitian if
\be
&&\beta^{\dag}=\beta,\;\;\;\ga=\al^{\dag}
\e{b2}
which we require. Eq.\r{b1} is easily solved algebraically. We find
\be
\frac{v^1}{v}\sin\frac{v}{2}=\sinh
{\frac{\beta}{2}}\cos{\frac{\alpha-\gamma}{2}}
\e{b3}
\be
\frac{v^2}{v}\sin\frac{v}{2}=i\sinh
{\frac{\beta}{2}}\sin{\frac{\alpha-\gamma}{2}}
\e{b4}
\be
\frac{v^3}{v}\sin\frac{v}{2}=\cosh
{\frac{\beta}{2}}\sin{\frac{\alpha+\gamma}{2}}
\e{b5}
\be
\cos\frac{v}{2}=\cosh{\frac{\beta}{2}}\cos{\frac{\alpha+\gamma}{2}}.
\e{b6}
All these relations are consistent with hermitian $v^a$ and \r{b2}. For
the measure we have formally
\be
&&d^3v\frac{2(1-\cos v)}{v^2}=id\al d\beta d\ga\sinh\beta
\e{b7} \\ \\
{\bf Case II}:
\be
&&e^{\psi_av_a}=e^{\al\psi_3}e^{\beta\psi_2}e^{\ga\psi_3}
\e{b8}
which requires \r{b2} and
\be
\frac{v^1}{v}\sin\frac{v}{2}=i\sin{\frac{\beta}{2}}\sin{\frac{\alpha-\gamma}{2}}
\e{b9}
\be
\frac{v^2}{v}\sin\frac{v}{2}=\sin{\frac{\beta}{2}}\cos{\frac{\alpha-\gamma}{2}}
\e{b10}
\be
\frac{v^3}{v}\sin\frac{v}{2}=\cos{\frac{\beta}{2}}\sin{\frac{\alpha+\gamma}{2}}
\e{b11}
\be
\cos\frac{v}{2}=\cos{\frac{\beta}{2}}\cos{\frac{\alpha+\gamma}{2}}.
\e{b12}
 For the
measure we find the formal relation
\be
&&d^3v\frac{2(1-\cos v)}{v^2}=-id\al d\beta d\ga\sin\beta
\e{b13}
\\ \\
 {\bf Case III}:
\be
&&e^{\psi_av_a}=e^{\al\psi_1}e^{\beta\psi_3}e^{\ga\psi_1}
\e{b14}
which requires \r{b2} and
\be
\frac{v^1}{v}\sin\frac{v}{2}=\cos{\frac{\beta}{2}}
\sinh{\frac{\alpha+\gamma}{2}}
\e{b15}
\be
\frac{v^2}{v}\sin\frac{v}{2}=-i\sin{\frac{\beta}{2}}
\sinh{\frac{\alpha-\gamma}{2}}
\e{b16}
\be
\frac{v^3}{v}\sin\frac{v}{2}=\sin{\frac{\beta}{2}}
\cosh{\frac{\alpha-\gamma}{2}}
\e{b17}
\be
\cos\frac{v}{2}=\cos{\frac{\beta}{2}}\cosh{\frac{\alpha+\gamma}{2}}.
\e{b18}
 For the
measure we find
\be
&&d^3v\frac{2(1-\cos v)}{v^2}=id\al d\beta d\ga\sin\beta
\e{b19}
\\ \\ {\bf Case IV}:
\be
&&e^{\psi_av_a}=e^{\al \phi_1}e^{\beta \phi_2}e^{\ga \phi_1}
\e{b20}
where
\be
&&\phi_1=\frac1{\sqrt2}(\psi_1+\psi_2),\;\;\;
\phi_2=\frac1{\sqrt2}(\psi_1-\psi_2),\;\;\;\phi_3=\psi_3
\e{b21}
Eq.\r{b20} requires \r{b2} and
\be
\frac{v^1+v^2}v\sin\frac v2
=\frac{\sqrt2}{2}(\alpha+\gamma+\frac12\alpha\beta\gamma)
\e{b22}
\be
\frac{v^1-v^2}v\sin\frac v2=\frac{\sqrt2}{2}\beta,
\e{b23}
\be
i\frac{v^3}v\sin\frac v2=\frac14\beta(\alpha-\gamma),
\e{b24}
\be
\cos\frac v2=\frac14(4+\beta(\alpha+\gamma))
\e{b25} For the measure we find
\be
&&d^3v\frac{2(1-\cos v)}{v^2}=-id\al d\beta d\ga \beta
\e{b26}
\\ \\
{\bf Case V}:
\be
&&e^{\psi_av_a}=e^{\al \phi_2}e^{\beta \phi_1}e^{\ga \phi_2}
\e{b27}
which requires \r{b2} and
\be
\frac{v^1+v^2}v\sin\frac v2=\frac{\sqrt2}{2}\beta
\e{b28}
\be
\frac{v^1-v^2}v\sin\frac v2
=\frac{\sqrt2}{2}(\alpha+\gamma+\frac12\alpha\beta\gamma),
\e{b29}
\be
i\frac{v^3}v\sin\frac v2=-\frac14\beta(\alpha-\gamma),
\e{b30}
and \r{b25}.
For the measure we find the relation \r{b26} here as well.


\begin{thebibliography}{Fierz}

\bibitem{Simple}R. Marnelius, \ {\sl Nucl. Phys.}\ {\bf B395}, 647 (1993)

\bibitem{BM}I. A. Batalin and R. Marnelius,\  {\em Solving general
gauge theories on inner product
spaces.} ITP-G\"oteborg preprint 94-32 (1994) (hep-th/9501004)

\bibitem{Propa} R. Marnelius, \ {\sl Nucl.
        Phys.}\ {\bf B418}, 353 (1994)




\bibitem{BV}I. A. Batalin and G. A. Vilkovisky, \
{\sl Phys. Lett.}
\ {\bf 69B},
309 (1977)

\bibitem{Bi}R. Marnelius, \
{\sl Nucl. Phys.}\ {\bf B370}, 165 (1992)


\bibitem{Gauge} R. Marnelius, \ {\sl Nucl.
        Phys.}\ {\bf B412}, 817 (1994)






\bibitem{Fermi} R. Marnelius, \ {\sl Int. J. Mod. Phys.} \  {\bf A5}, 329
(1990)



\bibitem{Pauli}W. Pauli, \  {\sl Rev. Mod. Phys.}\ {\bf 15}, 175
(1943)

\bibitem{Gen}R. Marnelius,
\  {\sl Nucl. Phys.}\ {\bf B391}, \ 621  \ (1993)

\bibitem{MA}R. Marnelius, \ {\sl Phys. Rev.}\ {\bf D20}, 2091 (1979)

\bibitem{MN}R. Marnelius and B. Nilsson, \ {\sl Phys. Rev.}\ {\bf D22}, 830
(1980)


\bibitem{MAN}R. Marnelius and B. Nilsson,
\  {\sl  Phys. Rev.}\ {\bf D20}, \ 839  \ (1979)




\end{thebibliography}
\end{document}